\documentclass{iopart}
\usepackage{epsf}
\begin{document}
\title{Tricritical wedge filling transitions with short-ranged forces}
\author{J. M. Romero-Enrique$^\dag$ and A. O. Parry$^\ddag$}
\address
{\dag\ Departamento de F\'{\i}sica At\'omica, Molecular y
Nuclear, Area de F\'{\i}sica Te\'orica, Universidad de Sevilla,
Apartado de Correos 1065, 41080 Sevilla, Spain}
\address{\ddag\ Department of Mathematics, Imperial College 180 Queen's Gate,
London SW7 2BZ, United Kingdom}
\begin{abstract}
We show that the 3D wedge filling transition in the presence of short-ranged
interactions can be first-order or second order depending on the strength
of the line tension associated with to the wedge bottom. This fact implies the
existence of a tricritical point characterized by a short-distance expansion
which differs from the usual continuous filling transition. Our analysis is 
based on an effective one-dimensional model for the 3D wedge filling which 
arises from the identification of the breather modes as the only relevant 
interfacial fluctuations. From such analysis we find a 
correspondence between continuous 3D filling at bulk 
coexistence and 2D wetting transitions with random-bond disorder. 

\end{abstract}
\maketitle

Fluid adsorption in micropatterned and sculpted geometries has become the 
subject of intense study over the last decade. Highly impressive 
technological advances which allow the tailoring of micro-patterned and
structured solid surfaces on the nanometer to micrometer scale 
\cite{herminghaus} are a landmark in the development of the
emerging microfluidic industry \cite{microfluid} which aims at minituarizing
chemical synthesis plants or biological analysis equipment in much the
same way the silicon chip brought about the electronics revolution.
However, the theoretical understanding of this phenomenon is far from being
complete. Recent studies of filling transitions for fluids in 3D
wedges show that interfacial fluctuations are greatly enhanced
compared with wetting at flat substrates \cite{Parry,Parry2}.
The control of such enhanced interfacial fluctuations is crucial 
for the effectiveness of the microfluidic devices. Fortunately,
there are simple theoretical approaches which take into account these effects.
For example effective Hamiltonian predictions for the critical exponents at 
continuous (critical) wedge filling with short-ranged forces have been 
confirmed in large scale Ising model simulation studies \cite{Milchev}. 
Similar experimental verification of the predicted geometry-dominated 
adsorption isotherms at complete wedge filling \cite{Bruschi} raise hopes
that the filling transition itself and related fluctuation effects will be
observable in the laboratory. We further develop the theory of wedge filling
in this paper, focussing on the emergence of a new type of continuous filling:
tricritical filling. 

\begin{figure}
\begin{center}
\epsfxsize=9cm
\epsfbox{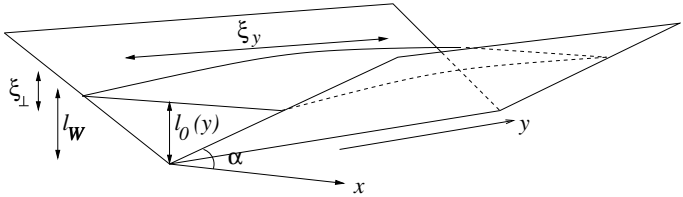}
\end{center}
\caption{Schematic illustration of a typical interfacial configuration
and relevant lengthscales for a fluid adsorption in a 3D wedge.
\label{fig1}}
\end{figure}

First we briefly review the fluctuation theory of 3D wedge filling. Consider 
the interface between a bulk vapour at temperature $T$ and saturation pressure 
with a 3D wedge characterised by a tilt angle $\alpha$.
Macroscopic arguments dictate that the wedge is partially filled by liquid if 
the contact angle $\theta>\alpha$ and completely filled if $\theta< \alpha$ 
\cite{Concus}. The filling transition refers to the change from microscopic to 
macroscopic liquid adsorption as $T\to T_f$, at which $\theta(T_f)=\alpha$, and
may be first-order or continuous (critical filling). Both of these transitions 
can be viewed as the unbinding of the liquid-vapour interface from the wedge 
bottom. Characteristic length scales are the mean interfacial height above the 
wedge bottom $l_W$, the roughness $\xi_\perp$ and the longitudinal correlation 
length $\xi_y$, measuring fluctuations along the wedge (see \Fref{fig1}). 
The relevant scaling fields at critical filling are $\theta-\alpha$ and the 
bulk ordering field $h$ (which is proportional to the pressure difference with
the saturation value). At coexistence ($h=0$) we define critical exponents by 
$l_W\sim (\theta-\alpha)^{-\beta_W}$ and $\xi_y \sim (\theta-\alpha)^{-\nu_y}$.
The roughness can be related to $\xi_y$ by the scaling relationship 
$\xi_\perp\sim \xi_y^{\zeta_W}$, where $\zeta_W$ is the wedge wandering 
exponent. For short-ranged forces, $\zeta_W=1/3$.

For shallow wedges, i.e. $\alpha\ll 1$, the free energy of 
an interfacial configuration can be modelled by an effective Hamiltonian
based on the capillary wave model for wetting of planar substrates 
\cite{Rejmer}. However, an analysis of this model \cite{Parry} shows
that the liquid-vapour interface across the wedge is aproximately flat and
soft-mode fluctuations arise from local translations in the height of
the filled region along the wedge. These \emph{breather modes}
are the only relevant fluctuations in the continuous filling phenomena, and
can be taken into account by the following effective Hamiltonian \cite{Parry}
\begin{equation}
{\mathcal H}_W[l_0]=\int d y\left\{\frac{\Sigma l_0}{\alpha}
\left(\frac{dl_0}{dy}\right)^2
+V_W(l_0)\right\}
\label{heff2}
\end{equation}
where $l_0(y)$ is the local height of the interface at position $y$ along
the wedge bottom and $\Sigma$ is the liquid-vapor surface tension. Note that
the effective bending term resisting fluctuations along the wedge is
proportional to the local interfacial height. 
The effective binding potential $V_W(l_0)$ for $l_0\gg l_\pi$, where
$l_\pi$ is the mean wetting film thickness for a planar substrate, is given
(up to irrelevant additive constants) by \cite{Parry}:
\begin{equation}
V_W(l_0)\approx \frac{h(l_0-l_\pi)^2}{\alpha}+\frac{\Sigma(\theta^2-\alpha^2) 
l_0} {\alpha}+ \int_{-(l_0-l_\pi)/\alpha}^{(l_0-l_\pi)/\alpha} dx 
W(l_0-\alpha |x|)
\label{effbinding}
\end{equation}
where $W(l)$ is the binding potential between the gas-liquid interface 
and a planar substrate. Note that the mean field result for $l_W$ is recovered 
by minimizing $V_W(l_0)$ for $l_0>l_\pi$, which is an important check
on the self-consistency of the method \cite{Parry}. In addition to a hard
wall repulsion for $l_0<0$, the potential $V_W(l_0)$ contains a short-ranged
attraction which may be modified by micropatterning a stripe along the
wedge bottom, so as to weaken the local wall-fluid substrate and therefore
strengthen the interfacial binding. This observation will be crucial for the 
existence of tricritical filling, since with this decoration it may
be possible to bind the interface to the wedge bottom at the filling boundary
$\theta=\alpha$ and $h=0$. For latter convenience, hereafter we will set $h=0$ 
in our discussion of continuous filling.

The quasi-one-dimensional nature of the effective Hamiltonian \Eref{heff2} 
allows us to use the transfer-matrix formalism. In the continuum limit
the partition function is defined as a path integral \cite{Burkhardt} (setting
$k_B T=1$ for convenience)
\begin{equation}
Z[l_b,l_a,Y]=\int {\mathcal D}l_0 \exp(-{\mathcal H}_W[l_0])
\label{partfunc}
\end{equation}
where $Y$ is the wedge length and $l_a$ and $l_b$ are the endpoint heights.
The position-dependent stiffness introduces some ambiguity in the
definition of the path integral. This problem was already pointed out in 
Ref. \cite{Bednorz} and is related to the well-known ordering
problem in the quantization of classical Hamiltonians with
position-dependent masses. Similar issues also arise in solid state physics 
\cite{Thomsen}. Borrowing from the methods used to overcome these difficulties
we use the following definition
\begin{equation}
Z[l_b,l_a,Y]=\lim_{N\to \infty} \int dl_1\ldots dl_{N-1} \prod_{j=1}^N
K(l_j,l_{j-1},Y/N)\label{partfunc2}
\end{equation}
where $l_0\equiv l_a$ and $l_N\equiv l_b$, and $K(l,l',y)$ is defined as:
\begin{equation}
K(l,l',y)=\sqrt{\frac{\Sigma\sqrt{ll'}}{\alpha \pi y}}
\exp\left(-\frac{\Sigma \sqrt{ll'}}{\alpha y}(l-l')^2-
y V_W(l)\right)
\label{partfunc3}
\end{equation}
In the continuum limit the partition function becomes
\begin{equation}
Z(l_b,l_a,Y)=\sum_n \psi_n(l_b)\psi_n^*(l_a)
\textrm{e}^{-E_n Y}
\label{partfunc4}
\end{equation}
where the complete orthonormal set satisfy
\begin{equation}
\Bigg(-\frac{\alpha}{4\Sigma}\frac{\partial}
{\partial l} \left[\frac{1}{l}\frac{\partial}{\partial l}\right]
+
V_W(l)-\frac{3\alpha}{16\Sigma l^3}\Bigg)\psi=E\psi
\label{schrodinger1}
\end{equation}
In the thermodynamic limit $Y\to \infty$ we obtain the probability
distribution function (PDF) for the midpoint interfacial height $P_W(l_0)=
|\psi_0(l_0)|^2$, the wedge excess free energy $f_W=E_0$ and the longitudinal 
correlation length $\xi_y=1/(E_1-E_0)$. At this point we must remark that
\emph{any} definition of the path integral which is invariant upon exchanging
$l_a$ and $l_b$ leads to an Schr\"odinger equation similar to 
\Eref{schrodinger1} but with a different coefficient for the extra $1/l^3$ term
in the effective binding potential \cite{Bednorz}.

\begin{figure}
\begin{center}
\epsfxsize=9cm
\epsfbox{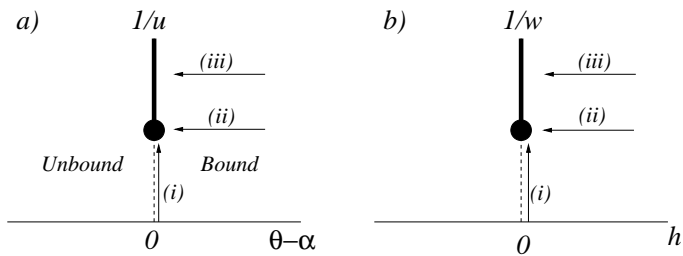}
\end{center}
\caption{Phase diagrams for (a) filling and (b) wetting transitions. The thick 
and dashed lines in both diagrams correspond to continuous and first-order
boundaries between bound and unbound interfacial states, respectively.
The arrows show representative paths along which continuous unbinding
occur: (i) and (ii) for tricritical filling (critical wetting) and (iii) for 
critical filling (complete wetting), respectively.  
The filled circles represent the tricritical filling and critical 
wetting points, respectively. See text for explanation.\label{fig2}}
\end{figure}

The change of variables $\lambda=\sqrt{8\Sigma/\alpha}l^{3/2}/3$ and 
$\psi(l)=(2\Sigma l/\alpha)^{1/4} \phi(\lambda(l))$ \cite{Yu} transforms the
\Eref{schrodinger1} to:
\begin{equation}
-\frac{1}{2}\frac{d^2 \phi(\lambda)}{d\lambda^2} + \left(V_W[l(\lambda)]-
\frac{5}{72 \lambda^2}\right)\phi(\lambda)=E \phi(\lambda)
\label{schrodinger2}
\end{equation} 
with $l(\lambda)=\left(3\lambda/\sqrt{8\Sigma/\alpha}\right)^{2/3}$. In 
general, there will be an interfacial bound state at bulk coexistence
for $\theta=\alpha$ if the strength of the small $l_0$ attraction between the 
gas-liquid interface and the substrates, which we will denote as $u$, is 
greater than some value $u_c$. Consequently, the filling transition is 
first-order if $u>u_c$ and critical if $u<u_c$. Tricritical 
filling is observed when $u-u_c$ emerges as a new relevant field (in the
renormalization-group sense). 
If $W(l)\sim -a/l^p+b/l^q$, different scenarios may arise as the range
of the binding potential is varied. In particular, for $p>4$ the long-range
behaviour of $V_W$ is dominated by the $1/l^3$ term for $\theta=\alpha$, so 
the filling phenomena are fluctuation-dominated. This finding is consistent 
with the existence of two different fluctuation regimes for critical filling: 
mean-field if $p<4$ and fluctuation-dominated regime if $p>4$ \cite{Parry}. 
In the fluctuation-dominated regime, examination of \Eref{schrodinger2} shows 
that there is an analogy between 3D continuous filling and 2D continuous 
wetting (see \Fref{fig2}), where the role at wetting of the bulk ordering 
field $h$ and the potential strength $w$ are played by $\theta-\alpha$ 
and $u$ for filling phenomena, respectively. In particular, 3D 
tricritical (critical) filling is analogous to 2D critical (complete) wetting, 
respectively. Different critical exponents which characterize the divergence of 
length scales can be defined. In addition to the critical filling critical 
exponents $\beta_W$ and $\nu_y$ defined along route (ii) in \Fref{fig2}(a) 
(see above), we can define new critical exponents for tricritical filling at 
$\theta=\alpha$ (route (i) in \Fref{fig2}(a)) as:
\begin{equation}
l_w\sim (u-u_c)^{-\beta_W^*} \ ,\ \xi_y\sim (u-u_c)^{-\nu_y^*}
\label{critexp1}
\end{equation}
and $\xi_\perp\sim \xi_y^{\zeta_W^*}$, where $\zeta_W^*$ is the tricritical
wandering exponent which in general may be different from $\zeta_W$ (in 
contrast with the wetting case). More generally, in the vicinity of the 
tricritical point we anticipate scaling e.g. $\xi_y\sim |u-u_c|^{-\nu_y^*} 
\Lambda \left[(\theta-\alpha)|u-u_c|^{-\Delta^*}\right]$ with the gap exponent
$\Delta^*$. Thus along route (ii) $\xi_y\sim (\theta-\alpha)^{-\nu_y^*/
\Delta^*}$.

We focus now on the case of short-ranged forces as the prototype of the
fluctuation-dominated regime. In addition to the hard-wall condition, 
$V_W(l_0)$ can be modelled as a contact-like attraction with strength $u$. 
An analysis of \Eref{schrodinger1} for $l_0\to 0$ shows that the short-distance
expansion of the PDF is either $P_W \sim l_0$ or $P_W \sim l_0^3$. We 
anticipate that the former corresponds to tricritical behaviour and the latter 
to critical filling. It is remarkable that thermodynamic consistency at 
critical filling is ensured as the local density at the wedge bottom is 
non-singular, i.e. $\rho_w(0)-\rho_l \sim T-T_f$, where $\rho_l$ is the bulk 
liquid density \cite{Parry3}. This property is only obtained if the partition 
function is defined by \Eref{partfunc2} and \Eref{partfunc3}. Thus, the 
ambiguity in its definition can be removed by imposing this regularity 
condition on the short-distance expansion of the interfacial height PDF. 

We report now our explicit results (details will be presented elsewhere).
Along route (i) we find that there is only one bound solution to 
\Eref{schrodinger1} for $u>u_c\approx 1.358$ with $E_0\propto (u-u_c)^3$ and 
associated PDF
\begin{equation}
P_W(l_0)=\frac{6\sqrt{3}\pi}{\xi_u}\frac{l_0}{\xi_u}
\left[\textrm{Ai}\left(\frac{l_0}{\xi_u}\right)\right]^2
\label{PDF}
\end{equation}
where $\textrm{Ai}(x)$ is the Airy function and in the scaling limit 
$\xi_u \sim |u-u_c|^{-1}$. Thus $l_W\sim \xi_\perp \propto (u-u_c)^{-1}$ 
and $\xi_y\propto (u-u_c)^{-3}$ identifying $\beta_W^*=1$, $\nu_y^*=3$ and 
obtaining $\zeta_W^*=\zeta_W=1/3$. As predicted, the short-distance behaviour 
of the PDF is linear with $l_0$. 

\begin{figure}
\begin{center}
\epsfxsize=9cm
\epsfbox{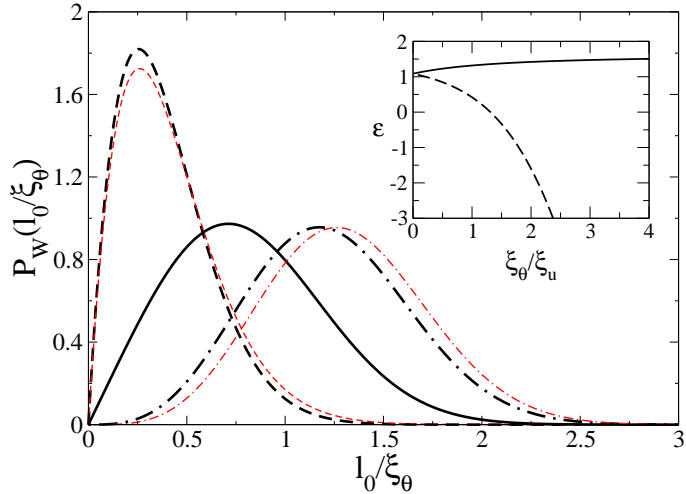}
\end{center}
\caption{Plot of the scaled PDF for $\epsilon=-1.5$ (thick dashed line), 
and along routes (ii) and (iii) in \Fref{fig2}(a), i.e. for $\epsilon\approx 
1.086$ (thick continuous line) and $\epsilon\approx 1.639$ (thick
dot-dashed line), respectively. For comparison, the PDF from \Eref{PDF} with 
$\xi_u\approx 1.968\xi_\theta$ (which corresponds to $\epsilon=-1.5$, see 
inset) is also plotted (thin dashed line). Finally, the scaled PDF 
obtained in Ref. \cite{Bednorz} is also shown (thin dot-dashed line). 
Inset: Plot of $\epsilon$ as a function of $\xi_\theta/\xi_u$ for $u<u_c$ 
(continuous line) and $u>u_c$ (dashed line).\label{fig3}}
\end{figure}

On the other hand, the scaling of the PDF for $\theta>\alpha$ is given by
(see also \Fref{fig3}):
\begin{eqnarray}
P_W(l_0) \propto l_0 \exp\left[
\frac{2 \epsilon l_0}{\xi_\theta} - \frac{2l_0^2}{\xi_\theta^2}
\right]
H_\nu^2\left(\sqrt{2}\frac{l_0}{\xi_\theta}
-\frac{\epsilon}{\sqrt{2}}\right)
\label{PDF2}
\end{eqnarray}
where $\xi_\theta=\Sigma^{-1/2}[(\theta/\alpha)^2-1]^{-1/4}$,
$\epsilon=\Sigma E_0 \xi_\theta^3/\alpha$, $\nu=\epsilon^2/4-1/2$ 
and $H_\nu(x)$ is the Hermite function \cite{Lebedev}. The value of 
$\epsilon$ is obtained as the smallest solution of the following
equation:
\begin{equation}
\pm \frac{\Gamma\left[-\frac{1}{3}\right]3^{-2/3}}{\Gamma\left[\frac{1}{3}
\right]}\frac{\xi_\theta}{\xi_u}=\epsilon + \left(\frac{
\epsilon^2}{\sqrt{2}}-\sqrt{2}\right)\frac {H_{\frac{\epsilon^2}{4}-
\frac{3}{2}}\left(-\frac{\epsilon}{\sqrt{2}}\right)}{H_{\frac{
\epsilon^2}{4}-\frac{1}{2}}\left(-\frac{\epsilon}{\sqrt{2}}\right)}
\label{groundstate2}
\end{equation}
where the positive (negative) sign corresponds to $u>u_c$ ($u<u_c$),
respectively. The inset of \Fref{fig3} plots the solution of this equation.
As anticipated, note that scaling is obeyed in the vicinity of the tricritical
point as the wedge excess free energy $f_W\sim 
\xi_\theta^{-3} F_\pm(\xi_\theta/\xi_u)$. For $u>u_c$ and $\xi_\theta/\xi_u\gg 
1$, we have checked numerically that the PDF \eref{PDF2} converges to the 
expression given by \Eref{PDF} for the corresponding value of $\xi_u$ given
by \Eref{groundstate2}. 
Thus, the interface remains bound to the substrate when $\theta\to \alpha$,
in agreement to the first-order character of the filling transition.
The thermodynamic path (ii) to the tricritical point corresponds to $\xi_u\to 
\infty$, which corresponds to $\epsilon\approx 1.086$. Thus along this route 
$l_W\sim \xi_\perp \propto (\theta-\alpha)^{-1/4}$ similar to critical filling.
From analysis of the spectrum it is also possible to show that $\xi_y\propto 
(\theta-\alpha)^{-3/4}$, so the tricritical gap exponent $\Delta^*=4$. 
Finally, for thermodynamic paths (iii) far from the tricritical 
point, i.e. $\xi_u \to 0$, we have found that the scaling of the PDF is of the 
form shown in \Eref{PDF2} with $\epsilon \approx 1.639$. Note that $P_W\sim 
l_0^3$ as $l_0\to 0$, in agreement with our previous statement. 
This condition is not fullfilled by the solution presented in Ref. 
\cite{Bednorz}, although globally it does not differ too much from our exact
solution (see \Fref{fig3}).

We finish by mentioning a remarkable connection for short-ranged forces 
between 3D wedge filling and 2D wetting with random-bond disorder 
\cite{preprint}. The critical exponents corresponding to tricritical and 
critical wedge filling can be obtained from generalized random-walk 
methods \cite{Fisher} in terms of the wedge wandering exponent $\zeta_W$.
In particular, they are found to have the same dependence of the critical 
exponents for critical and complete wetting, respectively, but in terms of
an effective 2D wandering exponent equal to $2\zeta_W$. For short-ranged forces 
($\zeta_W=1/3$), this implies that the set of critical exponents is the same 
as for 2D random-bond disorder \cite{Huse}. These predictions may certainly be 
tested in Ising model simulation studies and would be a stringent test of the 
theory of 3D wedge filling. 

\ack
J.M.R.-E. acknowledges financial support from the European Commission
under Contract MEIF-CT-2003-501042.

\end{document}